\documentclass[conference,final]{IEEEtran}

\usepackage{ifpdf}

\ifCLASSINFOpdf
  \usepackage[pdftex]{graphicx}
  \graphicspath{{../pdf/}{../jpeg/}}
  \DeclareGraphicsExtensions{.pdf,.jpeg,.png}
\else
  \usepackage[dvips]{graphicx}
  \graphicspath{{../eps/}}
  \DeclareGraphicsExtensions{.eps}
\fi

\usepackage{cite}
\usepackage{algorithm}
\usepackage{algorithmic}
\usepackage[cmex10]{amsmath}
\usepackage{amsthm}
\usepackage{amssymb}
\usepackage{balance}
\usepackage{eqparbox}
\usepackage{color}

\usepackage{url}

\newtheorem{proposition}{Proposition}
\newcommand{\bs}[1]{\boldsymbol{#1}}
\newcommand{\mc}[1]{\mathcal{#1}}

\newcommand{\bseq}{\begin{subequations}}
\newcommand{\eseq}{\end{subequations}}
\newcommand{\baln}{\begin{align}}
\newcommand{\ealn}{\end{align}}
\newcommand{\balnd}{\begin{aligned}}
\newcommand{\ealnd}{\end{aligned}}
\newcommand{\beq}{\begin{equation}}
\newcommand{\eeq}{\end{equation}}
\newcommand{\beqn}{\begin{eqnarray}}
\newcommand{\eeqn}{\end{eqnarray}}
\newcommand{\beqno}{\begin{eqnarray*}}
\newcommand{\eeqno}{\end{eqnarray*}}
\newcommand{\bma}{\begin{displaymath}}
\newcommand{\ema}{\end{displaymath}}
\newcommand{\bnu}{\begin{enumerate}}
\newcommand{\enu}{\end{enumerate}}
\newcommand{\bce}{\begin{center}}
\newcommand{\ece}{\end{center}}
\newcommand{\btb}{\begin{tabular}}
\newcommand{\etb}{\end{tabular}}
\newcommand{\ba}{\begin{array}}
\newcommand{\ea}{\end{array}}

\begin{document}

\title{Computation Offloading and Resource Allocation for Backhaul Limited Cooperative MEC Systems
}
%
%
\author{Phuong-Duy Nguyen${}^{\dagger}$, Vu Nguyen Ha${}^{\ddagger}$, and Long Bao Le${}^{\dagger}$ \\
${}^{\dagger}$INRS-EMT, University of Quebec, Montreal, Quebec, Canada; emails: \{phuongnguyen,le\}@emt.inrs.ca \\
${}^{\ddagger}$\'{E}cole Polytechnique de Montr\'{e}al, Montreal, Quebec, Canada; email: vu.ha-nguyen@polymtl.ca.
}

\maketitle

\begin{abstract}
In this paper, we jointly optimize computation offloading and resource allocation to
minimize the weighted sum of energy consumption of all mobile users in a backhaul limited cooperative MEC system with multiple fog servers. 
Considering the partial offloading strategy and TDMA transmission at each base station, the underlying optimization problem
with constraints on maximum task latency and limited computation resource at mobile users and fog servers is non-convex.
We propose to convexify the problem exploiting the relationship among some optimization variables from which
an optimal algorithm is proposed to solve the resulting problem. 
We then present numerical results to demonstrate the significant gains of our
proposed design compared to conventional designs without exploiting cooperation among fog servers and a greedy algorithm.
\end{abstract}


\section{Introduction}



Mobile Edge Computing (MEC) has been considered very potential in overcoming the limited computing resource constraint of mobile users (MUs)  
\cite{ETSI_2014} where MUs can offload partially or fully their computation load (CL) to the fog servers (FSs) at the network edge, which will be called partial and binary offloading, respectively. Such computation offloading (CO) can result in smaller energy consumption and/or better execution latency
 \cite{X_Lyu_NetWork18,TiLong2019}. However, CO typically requires transmission of some incurred data from MUs to the fog servers.
From the MU's perspective, CO can reduce the local CL but it also consumes extra energy for the data transmission. 
Therefore, joint CO decision and resource allocation (RA) becomes a major challenge in MEC systems which has been studied in
 several recent works \cite{C_Wang_TVT17,TiLong2019,Y_Wang_TCom16,K_Huang_TWC17,Mashayekhy_TCC15,G_Zhang_ICC18,X_He_VTC18,TXTran_TVT19}.
In particular, the papers \cite{C_Wang_TVT17,TiLong2019,Y_Wang_TCom16,K_Huang_TWC17} consider MEC systems in which one FS serves multiple MUs. Moreover, \cite{C_Wang_TVT17} designs a binary CO scheme to improve energy-efficiency considering radio interference while \cite{TiLong2019} proposes fair binary CO algorithms aiming to minimize the maximum weighted energy consumption. The partial CO strategy is considered in \cite{Y_Wang_TCom16,K_Huang_TWC17}. Both papers focus on minimizing the total energy consumption of all MUs with the delay
 constraint where \cite{Y_Wang_TCom16} investigates the offloading for the frequency division multiple access (FDMA)  while \cite{K_Huang_TWC17} studies the offloading for both OFDMA and TDMA scenarios.

A joint CO and RA design is recently investigated in multiple-FS MEC systems where a group of FSs is available to help the MUs reduce their CLs \cite{G_Zhang_ICC18,X_He_VTC18,TXTran_TVT19,oueis2014impact}.
In particular, \cite{G_Zhang_ICC18} studies the fair partial CO for the multiple-FS MEC system to minimize the task execution delay. 
In this work, a task from an MU is first partitioned into multiple subtasks which are offloaded independently to different
 FSs within their coverage using the corresponding wireless links.
In \cite{X_He_VTC18}, a binary CO strategy in the multi-fog system is proposed where CL from an MU can be executed at its FS or 
at one of its neighboring access points. 
Similarly, the authors in \cite{TXTran_TVT19} propose a joint CO and RA framework which maximizes the weighted sum of the reductions
 in task completion time and energy consumption. These papers do not consider cooperation among the FSs.
Considering FSs' cooperation, \cite{oueis2014impact} studies the multiple-cloud load balancing problem
for the system with limited backhaul capacity where minimization of the maximum CL of all cloud servers is the design objective.
However, optimization of the CO decisions is not addressed in this work.

Our current paper studies the joint CO and RA in a cooperative MEC system with multiple FSs 
where cooperative computation load balancing among different FSs is considered to efficiently serve the offloaded CLs. 
Specifically, an overloaded FS can offload its CL to neighboring FSs through the capacity limited backhaul links.
Our design aims at minimizing the weighted sum of energy consumption considering constraints on the maximum task execution
latency and computation resource (CR) at the MUs and FSs as well as the backhaul capacity. 
The design can be applied to any multi-FS MEC systems with arbitrary backhaul network topologies.
The optimization problem is formulated as a non-convex problem which is hard to solve due to the complex coupling among the variables. 
To overcome this challenge, we first study the relationship among the variables based on which the problem is convexified 
by using the penalty method. Then, a duality-based algorithm is proposed to solve the problem optimally.
For comparison purpose, a greedy algorithm is also discussed. 
Numerical results are then presented to illustrate the various insights and significant performance gains compared to conventional
design without exploiting cooperation among FSs and the greedy algorithm. 

\section{System Model and Problem Formulation}
\label{modelformu}


\begin{figure}
\centering
\includegraphics[width=0.6\columnwidth]{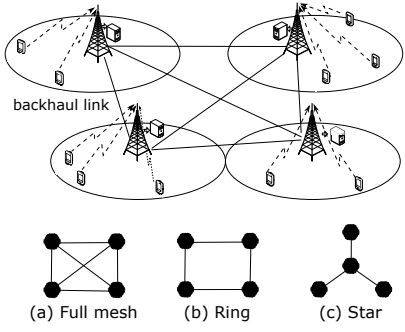}
\caption{Backhaul-limited cooperative MEC systems with different backhaul topologies.}
\label{fig-MultiFog}
\vspace{-5mm}
\end{figure}

Consider a multi-FS system consisting $S$ FSs deployed at $S$ one-antenna base stations (BS). 
Each BS is connected with a corresponding co-located FS which provide wireless communication and
 edge computation services to 
MUs inside its coverage. 
Let $\mathcal{S}$ denote the set of all FSs (or BSs), and $\mathcal{K}_s$ denote the set of $K_s$ MUs served by FS/BS $s$. Furthermore, 
it is assumed that each FS can offload its CL to other FSs in a FS set through the corresponding backhaul links. The helping
FS set for each FS can be determined from the underlying backhaul network topology, which is denoted as $\mathcal{L}_s$ for FS $s$.
For convenience, we also define $\mathcal{L}^{\sf{O}}_s$ as the set of FSs which can offload their CLs to FS $s$.

\subsection{Local-Computing and Offloading Model}

Assume that MU $k$ in $\mathcal{K}_s$ has to execute a computation task with $D_{k,s}$-bit input data within
the maximum latency of $\bar{T}_{k,s}$ seconds. 
The partial offloading strategy is considered in this paper.
In particular, $\ell_{k,s}$ bits of the incurred data are executed locally at the MU while the remaining $u_{k,s}$ bits will be 
offloaded to its FS for processing where $u_{k,s}=D_{k,s} - \ell_{k,s}$.
Let $f_{k,s}$ be the CPU clock speed of MU $k$ associated with BS $s$, which can be chosen in the range $(0,\bar{F}_{k,s}]$.
The local execution energy per CPU cycle can be expressed as $\rho_{k,s} = \alpha_{k,s} f_{k,s}^2$ \cite{W_Zhang_TWC13} where $\alpha_{k,s}$ represents the energy coefficient specified in the CPU model.
Denote $c_{k,s}$ as the number of CPU cycles required for executing $1$ data bit of MU $k$.  
Then, the energy and time consumed for local execution at MU $k$ can be expressed, respectively as
\beqn 
E^{\sf{Lo}}_{k,s} \!\! &{=}& \!\! c_{k,s} \alpha_{k,s} \ell_{k,s} f_{k,s}^2 = c_{k,s} \alpha_{k,s} (D_{k,s} -u_{k,s}) f_{k,s}^2, \label{E_Lo} \\
t^{\sf{Lo}}_{k,s} \!\! &{=}& \!\! {c_{k,s} \ell_{k,s}}/{f_{k,s}} = {c_{k,s} (D_{k,s} -u_{k,s})}/{f_{k,s}}.
\eeqn
Once MU $k$ decides to offload $u_{k,s}$ bits to FS $s$, it transmits these data bits  to BS $s$ over the wireless link.
The TDMA transmission strategy is employed in this paper.
In cell $s$ (corresponding to FS $s$), an interval of $T_s$ seconds is divided into $K_s$ time slots each of which is then assigned 
to one corresponding MU in the set $\mathcal{K}_s$ to support the computation offloading. 
Let $h_{k,s}$ be the channel gain between MU $k$ and BS $s$, which is assumed to be unchanged during the offloading duration;
 and $p_{k,s}$ denote the transmission power of MU $k$. 
To avoid the strong inter-cell interference, we assume that different frequency bands are allocated to different BSs \cite{Kosta_CST13}. 
Then, the data rate corresponding to the transmission from MU $k$ to its BS can be expressed as
\beq
r_{k,s} = W \log_2 \left( 1+{p_{k,s}h_{k,s}}/{\sigma^2} \right),
\eeq
where $W$ is the bandwidth and $\sigma^2$ is the noise power. 
Let $t_{k,s}$ denote the duration of the time slot assigned to MU $k$, which must satisfy 
$\sum \limits_{k \in \mathcal{K}_s} t_{k,s} = T_s$ due to the employed TDMA strategy.
Hence, the transmission power of MU $k$ in BS $s$ can be set so that $r_{k,s}={u_{k,s}}/{t_{k,s}}$. 
Thus, $p_{k,s}$ can be expressed as $p_{k,s} = \chi\left({u_{k,s}}/{t_{k,s}} \right)$ where $\chi(x)=\tilde{h}_{k,s}^{-1}\left( 2^{x/W}-1 \right)$, and $\tilde{h}_{k,s}={h_{k,s}}/{\sigma^2}$ \cite{K_Huang_TWC17}.
Then, the energy consumed by offloading MU $k$ can be expressed as
\beq \label{E_Off}
E^{\sf{Off}}_{k,s} =  t_{k,s} p_{k,s} = t_{k,s} \chi\left({u_{k,s}}/{t_{k,s}} \right).
\eeq
Combination of the results in \eqref{E_Lo} and \eqref{E_Off} yields the total energy consumed for MU $k$ at BS $s$ as
\beqn
E_{k,s} = E^{\sf{Lo}}_{k,s} + E^{\sf{Off}}_{k,s} = c_{k,s} \alpha_{k,s} \ell_{k,s} f_{k,s}^2 + t_{k,s} 
\chi\left({u_{k,s}}/{t_{k,s}} \right).
\eeqn

\subsection{Multi-Fog Cooperation Computing Model}

The offloaded CL from MU $k$ (corresponding to $u_{k,s}$ bits) can be executed at FS $s$ or be further offloaded to 
FSs in the set $\mathcal{L}_s$. Let
$u_{k,s,m}$ denote the number of bits, which is a part of $u_{k,s}$ bits, offloaded to FS $m$ ($m \in \mathcal{L}_s$) for 
processing, we have $\sum_{m \in \mathcal{L}_s} u_{k,s,m} = u_{k,s}$.
Assume that the allocated backhaul rate between FSs $s$ and $m$ is fixed at $d_{s,m}$, and the CPU clock speed that FS $m$ allocates for executing $u_{k,s,m}$ is $f_{k,s,m}$. 
Then, the time consumed for further offloading and executing  $u_{k,s,m}$ bits at BS $m$ can be given, respectively as
\beqn
t_{k,s,m}^{\sf{TF}} = {u_{k,s,m}}/{d_{s,m}} \text{ and }
t_{k,s,m}^{\sf{Fog}} = {u_{k,s,m} c_{k,s}}/{f_{k,s,m}}.
\eeqn
Note that $u_{k,s,s}$ stands for the number of bits processed directly at BS $s$; hence, $t_{k,s,s}^{\sf{TF}}$ should be set to be zero. To yield this required outcome, we set $d_{s,s}$ as $+\infty$, $s \in \mathcal{S}$. For many practical applications, the amount of
 data describing the computation outcome to be sent back to every MU, is usually much smaller than that due to the task offloading. Therefore, we omit the downlink transmission in this paper. Hence, the total time required for task execution of MU $k$ can be written as
\beq
T_{k,s} = \max\big(t^{\sf{Lo}}_{k,s}, T_s + \max \limits_{m \in \mathcal{L}_s}{(t_{k,s,m}^{\sf{TF}}+t_{k,s,m}^{\sf{Fog}})}\big).
\eeq

\subsection{Problem Formulation}
In this paper, we focus on minimizing the weighted sum energy consumption of all MUs and
 our design can be formulated in the following problem:
\begin{subequations} 
\begin{eqnarray} 
\hspace{-2cm} (\mathcal{P}_1)  &  \min\limits_{\Omega_1} & \sum \limits_{\forall (k,s)} \beta_{k,s} E_{k,s}  \\
&\text{s.t } &\ell_{k,s} + \sum_{m \in \mathcal{L}_s} u_{k,s,m} = D_{k,s}, \; \forall (k,s) , \label{C0}\\[-0 pt]
&& \sum \limits_{s \in \mathcal{L}^{\sf{O}}_m} \sum \limits_{k \in \mathcal{K}_s} f_{k,s,m} \leq \bar{F}_m, \; \forall m, \label{C1}\\[-0 pt]
&&f_{k,s} \leq \bar{F}_{k,s}, \; \forall (k,s), \label{C3}\\[-0 pt]
&& {c_{k,s} \ell_{k,s}}/{f_{k,s}}  \leq \bar{T}_{k,s}, \forall (k,s), \label{C_f_LO} \\
&& t_{k,s,m}^{\sf{TF}}+t_{k,s,m}^{\sf{Fog}} \!\! + \!\! \sum_{k \in \mathcal{K}_s} t_{k,s} \leq \bar{T}_{k,s}, \forall (k,s,m), \label{C_T_Fog}
\end{eqnarray}
\end{subequations}
where $\beta_{k,s}$'s are the weighted coefficients, $\Omega_1 = \{T_s,\ell_{k,s}, u_{k,s,m}, t_{k,s}, f_{k,s},f_{k,s,m}\}$'s 
denotes the set of all variables. Here, the constraints represent the backhaul capacity, computing resource, and 
different delay constraints where $\bar{F}_m$ is the maximum clock speed of FS $m$.

\section{Optimal Joint Computation Offloading and Resource Allocation}
\subsection{Penalty based Convex Transformation}

This section describes how to convexify problem $(\mathcal{P}_1)$ which is non-convex due to the non-linear 
term $\ell_{k,s} f_{k,s}^2$ in $E^{\sf{Lo}}_{k,s}$ and the non-convex constraint \eqref{C_T_Fog}.
Toward this end, we first study the relationship between the optimal value of $f_{k,s}$ with other variables in the following proposition.
\begin{proposition} \label{lmm1}
The optimal value of $f_{k,s}$ for any values of $\ell_{k,s}$ and $\bar{T}_{k,s}$ can be expressed as
\begin{equation} \label{f_local_opt}
f_{k,s}^{\star} = {\big(c_{k,s} \ell_{k,s}\big)}/{\bar{T}_{k,s}}. 
\end{equation}
\end{proposition}
\begin{IEEEproof}
As can be seen, constraint \eqref{C_f_LO} can provide the lower bound on $f_{k,s}$ as $\frac{c_{k,s} \ell_{k,s}}{\bar{T}_{k,s}} 
\leq f_{k,s}$. Besides, it can be verified from the local computing energy consumption given in \eqref{E_Lo} that the objective function of 
problem $(\mathcal{P}_1)$ is a monotonic increasing function of $f_{k,s}$.
Therefore, the optimal value of $f_{k,s}$ for any value of $\ell_{k,s}$ and $\bar{T}_{k,s}$ can be expressed in \eqref{f_local_opt}.
\end{IEEEproof}
The proof of Proposition~\ref{lmm1} also shows that the constraint \eqref{C3} will be violated if $\frac{c_{k,s} \ell_{k,s}}{\bar{T}_{k,s}} > \bar{F}_{k,s}$. Hence, one yields $\ell_{k,s} \leq \frac{\bar{T}_{k,s}\bar{F}_{k,s}}{c_{k,s}}$ which is equivalent to
\beq
D_{k,s} - {\bar{T}_{k,s}\bar{F}_{k,s}}/{c_{k,s}} \leq u_{k,s} \leq D_{k,s}.
\label{u_ks}
\eeq
By replacing $\ell_{k,s} = D_{k,s} - u_{k,s}$ and $f_{k,s} = {\big(c_{k,s} \ell_{k,s}\big)}/{\bar{T}_{k,s}}$ using Proposition~\ref{lmm1}, the variable set can be reduced to $\Omega_2=\Omega_1/\{f_{k,s},\ell_{k,s}\}\text{'s}$ and $E^{\sf{Lo}}_{k,s}$ can be rewritten as ${\alpha_{k,s} c_{k,s}^3 (D_{k,s} \!\! - \!\!u_{k,s})^3}/{\bar{T}_{k,s}^2}$  which is a convex function with respect to $u_{k,s} > 0$. However, problem $(\mathcal{P}_1)$ is still non-convex due to non-convex term $\frac{u_{k,s,m} c_{k,s}}{f_{k,s,m}}$ in constraint \eqref{C_T_Fog}.
To deal with this challenge, we introduce a new variable $a_{k,s,m}$ which satisfies $a_{k,s,m}^2 = u_{k,s,m}$ $\forall(k,s,m)$. 
Using these new variables, problem $(\mathcal{P}_1)$ can be rewritten as 
\begin{subequations} \label{p2}
\begin{eqnarray} 
\hspace{-6mm} (\mathcal{P}_2) \!\!\!\!  &  \!\!\!\! \min\limits_{\Omega_3} \!\!\!\! & \!\!\!\!\!\! \sum \limits_{\forall (k,s)} \!\!  \beta_{k,s} \!\! \left( \!\! \frac{\alpha_{k,s} c_{k,s}^3 (D_{k,s} \!\! - \!\!u_{k,s})^3}{\bar{T}_{k,s}^2} \! + \! t_{k,s} \chi \! \left(\frac{u_{k,s}}{t_{k,s}} \right)\!\! \right) \\[-0 pt]
\hspace{-6mm} & \text{s.t } & \text{constraints \eqref{C1}, \eqref{u_ks},} \nonumber\\[-0 pt]
\hspace{-6mm} && \!\! \sum_{m \in \mathcal{L}_s} a_{k,s,m}^2 = u_{k,s}, \forall (k,s), \label{C6}\\[-0 pt]
\hspace{-6mm} && \!\! \frac{a_{k,s,m}^2}{d_{s,m}} \!\! +  \!\! \frac{c_{k,s} a_{k,s,m}^2}{f_{k,s,m}} \!\! + \!\!\!\! \sum_{k \in \mathcal{K}_s} t_{k,s} \!\! \leq \!\! \bar{T}_{k,s}, \forall (k,s,m), \label{C7}
\end{eqnarray}
\end{subequations}
where $\Omega_3=\Omega_2 \cup \{a_{k,s,m}\}$'s. 
A can be seen, the left hand side of constraint \eqref{C7} is a convex function of $(a_{k,s,m},f_{k,s,m})$.
Thus, problem $(\mathcal{P}_2)$ is convex which can be solved optimally if one omits constraint \eqref{C6}.
Therefore, we employ the Penalty-Function method to solve problem $(\mathcal{P}_2) $ by dealing with the following
 problem
\begin{subequations} \label{p3}
\begin{eqnarray} 
\hspace{-2mm} (\mathcal{P}_3) & \!\!\! \min\limits_{\Omega_3} & \!\!\!\!\!\!  \sum_{\forall (k,s)} \!\!\! \Big[ \beta_{k,s} E_{k,s} {+} \rho \big( \!\!\! \sum \limits_{m \in \mathcal{L}_s} \!\!\! a_{k,s,m}^2 {-} u_{k,s} \big) ^2 \Big]  \\[-0 pt]
\hspace{-2mm} &\text{s.t } & \!\!\!\! \text{constraints \eqref{C1}, \eqref{u_ks}, \eqref{C7}.} 
\end{eqnarray}
\end{subequations}
In particular, the optimal solution of $(\mathcal{P}_2)$ can be obtained by repeatedly solving problem $(\mathcal{P}_3)$ and increasing $\rho$ until constraint \eqref{C6} holds. In the next section, we study how to tackle problem $(\mathcal{P}_3)$
and propose an optimal algorithm to solve problem $(\mathcal{P}_2)$.
Problem $(\mathcal{P}_3)$ is convex thanks to its convex-form objective function and constraints. Therefore, it can be solved by employing the standard duality method as presented in the following section.
\subsection{Duality-based Optimal Algorithm}
The Lagrangian of problem $(\mathcal{P}_3)$ is expressed in \eqref{lagran}\addtocounter{equation}{1} where $\bs{\mu}= \lbrace \mu_{m}\rbrace$'s, $\bs{\gamma}= \lbrace \gamma_{k,s,m}\rbrace$'s are the Lagrangian multipliers associated with the constraints of
 problem $(\mathcal{P}_3)$.
Then, the dual function $g(\bs{\mu},\bs{\lambda},\bs{\gamma})$ can be defined as 
\beq \label{dual-func}
g(\bs{\mu},\bs{\gamma}) = \inf_{ \Omega_3 } \mc{L}(\Omega_3 ,\bs{\mu},\bs{\gamma}) \text{ s.t. \eqref{u_ks}},
\eeq
and the dual problem can be stated as
\begin{eqnarray}\label{dual-prob}
\underset{\bs{\mu},\bs{\gamma}}{\mathrm{max}}\; g(\bs{\mu},\bs{\gamma}) \text{ s.t.} \; \mu_{m}, \gamma_{k,s,m} \geq 0, \forall (k,s,m). 
\end{eqnarray}
\newcounter{tempequationcounter}
\begin{figure*}[!t]
\normalsize
\setcounter{tempequationcounter}{\value{equation}}
\begin{IEEEeqnarray}{rCl}
\setcounter{equation}{17}
\label{lagran}
 \mathcal{L}(\Omega_3,\bs{\mu},\bs{\gamma}) &=& \sum_{ \forall (k,s)} \beta_{k,s} \left(\frac{\alpha_{k,s} c_{k,s}^3 (D_{k,s}-u_{k,s})^3}{\bar{T}_{k,s}^2} + t_{k,s} \tilde{h}_{k,s}^{-1} \left( 2^{ \frac{u_{k,s}}{t_{k,s}} . \frac{1}{W}}-1 \right) \right)+ \rho  \sum_{ \forall (k,s)} \left( \sum \limits_{m \in \mathcal{L}_s} a_{k,s,m}^2 - u_{k,s}\right) ^2 \nonumber\\ 
\hspace{-8mm} && + \sum_{m\in S}\mu_{m} \left( \sum \limits_{s \in \mathcal{L}^O_m} \sum \limits_{k \in \mathcal{K}_s} f_{k,s,m} - \bar{F}_m \right)   + \sum_{ \forall (k,s)}\sum \limits_{m \in \mathcal{L}_s} \gamma_{k,s,m} \left( \frac{a_{k,s,m}^2}{d_{s,m}} + \frac{c_{k,s} a_{k,s,m}^2}{f_{k,s,m}} + \sum \limits_{r \in \mathcal{K}_s} t_{r,s} - \bar{T}_{k,s} \right).
\end{IEEEeqnarray}
\setcounter{equation}{\value{tempequationcounter}}
\hrulefill
\vspace*{4pt}
\end{figure*}
From the KKT conditions \cite{Boyd2004}, the optimal solution $\Omega_3^{\star}=\lbrace u_{k,s}^{\star}, t_{k,s}^{\star}, a_{k,s,m}^{\star}, f_{k,s,m}^{\star} \rbrace $ of problem \eqref{dual-func} for given $\bs{\mu},\bs{\gamma}$ can be determined 
as stated in the following propositions.
\begin{proposition} \label{ppst2}
Let $z^{\star}_{k,s}=\frac{u^{\star}_{k,s}}{W t^{\star}_{k,s}}$. For given  $\bs{\mu},\bs{\gamma}$, the value of $z_{k,s}$ can be expressed as  
\beq \label{z_ks}
z^{\star}_{k,s} = {W_n \Big(\big( {\gamma_{k,s,m} \tilde{h}_{k,s}}/{\beta_{k,s}}-1 \big)/e + 1 \Big)}/{ln2},
\eeq
where $W_n(.)$ is the Lambert function \cite{Corless1996}.
\end{proposition}
\begin{IEEEproof}
Due to the space constraint, the proof is given briefly as follows.
Applying the KKT conditions, we first take the derivative of $\mathcal{L}(\Omega_3,\bs{\mu},\bs{\gamma})$ with respect to $t_{k,s}$, 
set it to zero, which results in an equation of $u_{k,s}/t_{k,s}$.
Substituting $z_{k,s}=\frac{u_{k,s}}{W t_{k,s}}$ into this equation yields a Lambert function
of $z_{k,s}$. Solving this equation as in \cite{Corless1996}, we can obtain $z^{\star}_{k,s}$ as given in \eqref{z_ks}. 
\end{IEEEproof}
\begin{proposition} \label{ppst3}
Let $A(z_{k,s}) = \frac{\sigma^2}{h_{k,s}} \left( 2^{ z_{k,s}}-1 \right)$ where $z_{k,s}=\frac{u_{k,s}}{W t_{k,s}}$.
For given $\bs{\mu},\bs{\gamma}$, the following value of $\Omega_3^{\star}$ satisfy the KKT conditions:
\beqn
\hspace{-8mm}&& a^{\star}_{k,s,m} = \sqrt{\frac{\mu_{m}}{\gamma_{k,s,m} c_{k,s}}} f^{\star}_{k,s,m},\;\; t^{\star}_{k,s} ={u^{\star}_{k,s}}/{(W z^{\star}_{k,s})},  \label{repres_at} \\
\hspace{-8mm}&& 
f^{\star}_{k,s,m} = \left[ \left(\frac{2 \rho \Delta_{k,s}}{\gamma_{k,s,m}} - \frac{1}{d_{s,m}} \right)^{-1} c_{k,s}\right] ^{+} \label{repres_f} \\
\hspace{-8mm}&& u^{\star}_{k,s} = \min \Big( \max\big( D_{k,s} - {\bar{T}_{k,s}\bar{F}_{k,s}}/{c_{k,s}} ,\Gamma_{k,s} \big), D_{k,s} \Big), \label{repres_kkt_u_a_delta} 
\eeqn
where $\Gamma_{k,s} = D_{k,s} - \sqrt{\frac{\bar{T}_{k,s}^2\left( A(z^{\star}_{k,s})+ 2 \rho \Delta_{k,s} \right)}{3\alpha_{k,s} \beta_{k,s} c_{k,s}^3}},$ and $\Delta_{k,s}$ is determined by solving the following equation
\beqn \label{repres_kkt_u_delta}
\hspace{-12mm}&& D_{k,s} - \sqrt{\frac{\bar{T}_{k,s}^2\left( A(z_{k,s})+ 2 \rho \Delta_{k,s} \right)}{3\alpha_{k,s} \beta_{k,s} c_{k,s}^3}} \nonumber \\
\hspace{-12mm}&& \;\;\;\;\;\;\;\;\;\;\;\; - \sum \limits_{m \in \mathcal{L}_s} \frac{ c_{k,s}\mu_{m}}{\gamma_{k,s,m}} \left(\frac{2 \rho \Delta_{k,s}}{\gamma_{k,s,m}} - \frac{1}{d_{s,m}} \right)^{-2}  = \Delta_{k,s}.
\eeqn
\end{proposition}
\begin{IEEEproof}
Due to the space constraint, the proof is given briefly as follows. First, we set
\beq \label{delta}
\Delta_{k,s} = u_{k,s} - \sum \limits_{m \in \mathcal{L}_s} a_{k,s,m}^2.
\eeq
Applying the KKT conditions, we first take the derivative of $\mathcal{L}(\Omega_3,\bs{\mu},\bs{\gamma})$ with respect to all variables in $\Omega_3$ and set the results equal to zeros to form a system of equations based on which the variables in $\Omega_3$ can be stated as a function of $\Delta_{k,s}$ as in \eqref{repres_at}-\eqref{repres_kkt_u_a_delta}. 
Substituting \eqref{repres_kkt_u_a_delta} and \eqref{repres_at} into \eqref{delta}  yields \eqref{repres_kkt_u_delta}. 
Then, $\Omega_3^{\star}$ hence can be expressed as in \eqref{repres_at}-\eqref{repres_kkt_u_a_delta} after obtaining $\Delta_{k,s}$ by solving \eqref{repres_kkt_u_delta}. Since $\Omega_3^{\star}$ are calculated from the KKT conditions, $\Omega_3^{\star}$ must satisfy these conditions.
\end{IEEEproof}

Thanks to Proposition~\ref{ppst2}-\ref{ppst3}, $g(\bs{\mu},\bs{\gamma})$ can be determined for given $\bs{\mu},\bs{\gamma}$. The remaining work is to solve the duality problem \eqref{dual-prob} which is considered in the following proposition.
\begin{proposition}\label{prop-4}
The dual function $g(\bs{\mu},\bs{\gamma})$ is a concave function and its
sub-gradient at $\mu_{m}$ is $ \sum \limits_{s \in \mathcal{L}^O_m} \sum \limits_{k \in \mathcal{K}_s} f_{k,s,m} - \bar{F}_m$, and at $\gamma_{k,s,m}$ is $\frac{a_{k,s,m}^2}{d_{s,m}} + \frac{c_{k,s} a_{k,s,m}^2}{f_{k,s,m}} + \sum \limits_{k \in \mathcal{K}_s} t_{k,s}  - \bar{T}_{k,s}$
where $\Omega_4^{\star}$ is obtained from the previous section. 
\end{proposition}
\begin{IEEEproof}
The dual function is a concave function by nature \cite{Boyd2004}.
The choice of the sub-gradient for $\mu_{m}$ and $\gamma_{k,s,m}$ is justified by the fact that $\bs{\mu}$ and $\gamma_{k,s,m}$ are the Lagrangian multipliers associated with constraints \eqref{C1} and \eqref{C7}, respectively. Details of this proof are omitted for brevity.
\end{IEEEproof}
Due to proposition \ref{prop-4}, the dual variables can be solved using the following iterative updates:
\beqn 
\mu_{m}^{(n+1)} &=&\mu_{m}^{(n)} + \delta_{m}^{(n)} \bigg( \sum \limits_{s \in \mathcal{L}^O_m} \sum \limits_{k \in \mathcal{K}_s} f_{k,s,m} - \bar{F}_m \bigg), \label{mu-update} \\
\gamma_{k,s,m}^{(n+1)} &=& \gamma_{k,s,m}^{(n)} + \alpha_{k,s,m}^{(n)} \times \nonumber\\
&& \!\!\!\!\!\!\!\! \bigg( \frac{a_{k,s,m}^2}{d_{s,m}} + \frac{c_{k,s} a_{k,s,m}^2}{f_{k,s,m}} + \sum \limits_{k \in \mathcal{K}_s} t_{k,s}  - \bar{T}_{k,s} \bigg), \label{gamma-update}
\eeqn
where  $\delta_{m}^{(n)}$ and $\alpha_{k,s,m}^{(n)}$ are suitable small step-sizes.
If $\delta_{m}^{(n)}, \alpha_{k,s,m}^{(n)} \overset{n\rightarrow\infty}\longrightarrow 0$,
the above sub-gradient based updates are guaranteed to converge to the optimal solution of problem \eqref{dual-prob}.
By iteratively updating $\left\lbrace \Omega_3^{\star}, \bs{\mu}, \bs{\gamma} \right\rbrace$, we can obtain the
optimal solution. The proposed solution approach is summarized in the Algorithm~\ref{alg:gms1} where $h(\Omega_3^{(n)})=\sum_{\forall (k,s)} |\Delta^{(n)}_{k,s}|$.

\begin{algorithm}[!t]
\footnotesize
\caption{\textsc{Duality-based Optimal Algorithm }}
\label{alg:gms1}
\begin{algorithmic}[1]
\STATE Initialization: Set $\bs{\mu}=\bs{0}$ for all $(k,s,m)$, choosing $\rho >0$, a tolerance $\varepsilon$, and set $n=0$.
\REPEAT
\REPEAT
\STATE Update $\Omega_3^{(n)}$ by following all steps in Propositions~\ref{ppst2}-\ref{ppst3}.
\STATE Update $\bs{\mu}^{(n+1)}$ and $\bs{\gamma}^{(n+1)}$ as in \eqref{mu-update}-\eqref{gamma-update}.
\STATE Update $n=n+1$.
\UNTIL Convergence.
\STATE Increasing $\rho$ by setting $\rho := M \rho$ ($M > 1$).
\UNTIL{$h(\Omega_3^{(n)}) \leq \varepsilon$}
\STATE {Output:} Obtain $\Omega_3^{opt} = \{ f_{k,s,m}^{opt},u_{k,s}^{opt}, a_{k,s,m}^{opt}\, t_{k,s}^{opt}\}$ and compute $E_{k,s}$ for each MU $k$ of BS $s$
\end{algorithmic}
\end{algorithm}

\section{Greedy Algorithm}
For a comparison purpose, a greedy algorithm is described in Algorithm~\ref{alg:gms2}.
This algorithm aims to perform load balancing among FSs. Specifically, we attempt to determine if one FS is overload and there exists a \textit{Ready-To-Help (RTH)} FS with available computing resource (CR) as follows.
First, we optimize the CO and RA at each BS/FS separately, from which the available CR of each FS is calculated as $\Delta f_s= \bar{F}_s-\sum_{\forall (k,m)} f_{k,m,s}, \forall s$. Then, the FS having the largest available CR is selected as \textit{RTH} FS $\hat{m}$. This \textit{RTH} FS will utilize all available CR to help the FS that has used all CR and its MUs spend the highest energy consumption, named as \textit{Asking-For-Help (AFH)} FS $s^{\star}$. 
Specifically, FS $s^{\star}$ is allowed to offload its CL to FS $\hat{m}$. 
The CPU clock speed that FS $\hat{m}$ assigns MUs in cell $s^{\star}$ can be determined as in \eqref{f_ksm_assigned}. Then, the offloaded data size $u_{k,s^{\star},\hat{m}}$ can be determined to achieve time balancing as follows:
\beq
\frac{c_{k,s^{\star}} (u_{k,s^{\star},s^{\star}} - u_{k,s^{\star},\hat{m}})}{f_{k,s^{\star},s^{\star}}} = \frac{u_{k,s^{\star},\hat{m}}}{d_{s^{\star},\hat{m}}} + \frac{c_{k,s^{\star}} u_{k,s^{\star},\hat{m}}}{f_{k,s^{\star},\hat{m}}},
\eeq
which yields $\eqref{u_ksm_assigned}$. Thanks to the load balancing between FSs $\hat{m}$ and $s^{\star}$ as in \textbf{Step 11} of Algorithm~\ref{alg:gms2}, the processing time at the FS for MUs in cell $s^{\star}$ is reduced by $\Delta T_{s^{\star}} = \min_{k \in \mathcal{K}_{s^{\star}}} (\frac{c_{k,s^{\star}} u_{k,s^{\star},\hat{m}}}{f_{k,s^{\star},s^{\star}}})$.
This time reduction can be exploited for increasing the offloading time of all MUs in cell $s^{\star}$. Hence, the new offloading duration can be defined as $T^{\prime}_{s^{\star}} = T_{s^{\star}} + \Delta T_{s^{\star}}$. Then, the join CO and RA design for cell $s^{\star}$ can be re-optimized by solving the following problem:
\begin{subequations} \label{p_one_cell}
\begin{eqnarray} 
(\mathcal{P}_{s^{\star}})  \!\! & \!\!  \underset{\Omega_s}{\min} \!\! & \!\!\!\! \sum_{k \in \mathcal{K}_{s^{\star}}} \!\!\!\! \beta_{k,s^{\star}} \!\! \left[ \frac{\alpha_{k,s^{\star}} c_{k,s^{\star}}^3 (D_{k,s^{\star}} - u_{k,s,s} - u_{k,s^{\star},\hat{m}})^3}{\bar{T}_{k,s^{\star}}^2} \right.   \nonumber \\
&& \;\;\;\;\;\;\;\;\; \left. + t_{k,s} \frac{\sigma^2}{h_{k,s^{\star}}} \left( 2^{ \frac{u_{k,s,s}+u_{k,s^{\star},\hat{m}}}{W t_{k,s}}}-1 \right) \right] \nonumber \\
&\text{s.t } & \!\!\!\! \text{\eqref{C1}, \eqref{u_ks}, \eqref{C_T_Fog} for cell $s^{\star}$ and } \sum_{k \in \mathcal{K}_{s^{\star}}} t_{k,s} \leq T^{\prime}_{s^{\star}}, \nonumber
\end{eqnarray}
\end{subequations}
where $\Omega_s = \lbrace u_{k,s,s}, f_{k,s,s}, t_{k,s} \rbrace_{k \in \mathcal{K}_{s^{\star}}}$.
Problem $(\mathcal{P}_{s^{\star}})$ is applied for one FS which can be solved by employing the algorithms in \cite{K_Huang_TWC17} or our proposed algorithm with low complexity.
Since the previous values of $\lbrace u_{k,s^{\star},s^{\star}}, f_{k,s^{\star},s^{\star}}, t_{k,s^{\star}} \rbrace_{k \in \mathcal{K}_{s^{\star}}}$ is a feasible solution of $(\mathcal{P}_{s^{\star}})$, solving $(\mathcal{P}_{s^{\star}})$ results in a new solution
 with smaller weighted sum of power consumption at cell $s^{\star}$.  \textbf{Step 4}-\textbf{Step 13} of Algorithm~\ref{alg:gms2} are repeated until there is no \textit{RTH} FS or \textit{AFH} FS.
\begin{algorithm}[!t]
\footnotesize
\caption{\textsc{Greedy Algorithm }}
\label{alg:gms2}
\algsetup{indent=1.5em}
\begin{algorithmic}[1]
\STATE Initialize $u_{k,s,m} = 0$ $\forall (k,s,m)$, $m \neq s$ and $T^{\prime}_s = \max_{k \in \mathcal{K}_s} \bar{T}_{k,s}$ $\forall s$, $\mathcal{S}^{\star}=\mathcal{S}$.
\STATE Solve $(\mathcal{P}_{s})$ for all cells to obtain optimum $\Omega_3^{\star}$.
\REPEAT
\STATE Calculate $\Delta f_s= \bar{F}_s-\sum\limits_{\forall (k,m)} f_{k,m,s}$.
\IF {$\Delta f_s > 0$ for all $s \in \mathcal{S}^{\star}$} 
\STATE \textbf{Stop} since there no bottleneck at FSs.
\ELSE
\STATE Retrieve $\hat{m} = \arg\max_{m \in \mathcal{S}} \Delta f_m$ as \textit{RTH} FS.
\STATE Update $\mathcal{S}^{\star}=\mathcal{S}/\{\hat{m}\}.$
\STATE Retrieve $s^{\star} = \arg\max_{s \in \mathcal{S}^{\star}} \sum_{k \in \mathcal{K}_s} \beta_{k,s} E_{k,s} \text{ s. t. } \Delta f_s = 0$.
\STATE Determine $f_{k,s^{\star},\hat{m}}$ and $u_{k,s^{\star},\hat{m}}$ as 
\beqn
f_{k,s^{\star},\hat{m}} \!\!\!\!\! & {=} & \!\!\!\!\! \Delta f_{\hat{m}} \frac{f_{k,s^{\star},s^{\star}}}{\sum_{k \in \mathcal{K}_{s^{\star}}} f_{k,s^{\star},s^{\star}}}, \label{f_ksm_assigned}\\
u_{k,s^{\star},\hat{m}} \!\!\!\!\! &{=}& \!\!\!\!\! u_{k,s^{\star},s^{\star}} \Big(\frac{f_{k,s^{\star},s^{\star}}}{c_{k,s^{\star}} d_{s^{\star},\hat{m}}} {+} \frac{f_{k,s^{\star},s^{\star}}}{f_{k,s^{\star},\hat{m}}} {+}1\Big)^{-1}.
\label{u_ksm_assigned}
\eeqn
\STATE Determine $T^{\prime}_{s^{\star}} = \sum \limits_{k \in \mathcal{K}_{s^{\star}}} t_{k,s^{\star}} + \min_{k \in \mathcal{K}_{s^{\star}}} (\frac{c_{k,s^{\star}} u_{k,s^{\star},\hat{m}}}{f_{k,s^{\star},s^{\star}}})$.
\STATE Solve $(\mathcal{P}_{s^{\star}})$ to update $\Omega_{s^{\star}}$.
\ENDIF
\UNTIL $\Delta f_s > 0$ for all $s \in \mathcal{S}^{\star}$.
\end{algorithmic}
\end{algorithm}

\section{Numerical Results} \label{sec:sim}

\begin{figure}[!t]
\centering
\includegraphics[width=70mm]{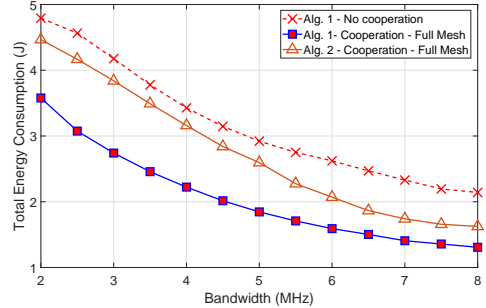}
\caption{Total energy consumption versus the frequency bandwidth $W$.}
\label{fig-Energy_W}
\vspace{-3mm}
\end{figure}

\begin{figure}[!t]
\centering
\includegraphics[width=70mm]{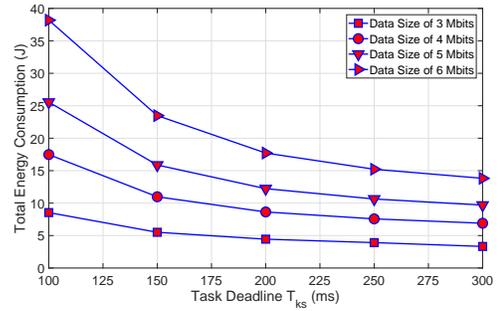}
\caption{Total energy consumption versus allowable task latency $T$.}
\label{fig-E_Time}
\vspace{-3mm}
\end{figure}

\begin{figure}[!t]
\centering
\includegraphics[width=70mm]{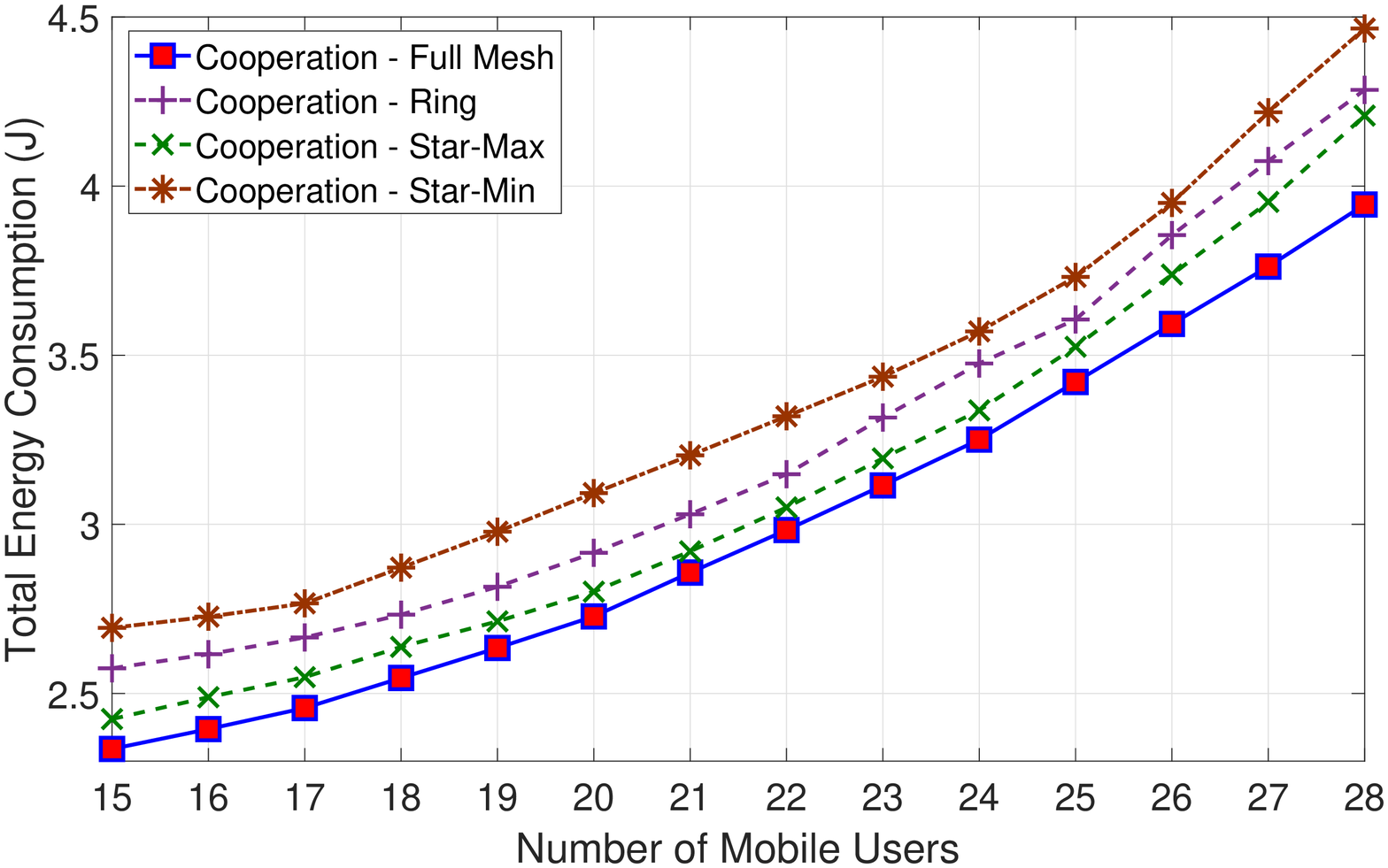}
\caption{Total energy consumption versus the number of MUs.}
\label{fig-E_MU}
\vspace{-3mm}
\end{figure}

\begin{figure}[!t]
\centering
\includegraphics[width=70mm]{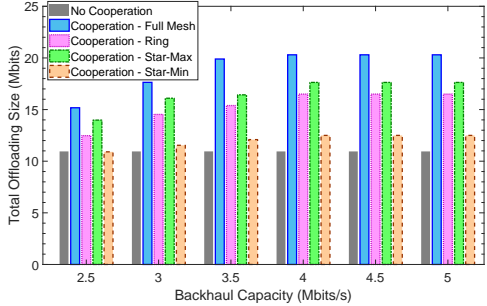}
\caption{Size of offloaded data versus the capacity of backhaul links.}
\label{fig-Offload}
\vspace{-3mm}
\end{figure}

The performance of the proposed algorithms for backhaul limited cooperative MEC system is investigated via numerical studies.
 We consider a simple 4-cell network where the distance between two nearest BSs is $400 \, m$ as illustrated in Fig.~\ref{fig-MultiFog}. 
In each cell, we randomly place 7 MUs (except for Fig.~\ref{fig-E_MU}) so that the distance from the cell center to every MU is in the range $[50 m,200 m]$.
The channel gains are generated by considering both Rayleigh fading and path loss which is modelled as 
$L_{k,s}^{k}=36.8 \mathsf{log}_{10}(y_{k,s}^k)+43.8+20\mathsf{log}_{10}(\frac{f_{\sf{freq}}}{5})$
where $y_{k,s}^k$ is the distance from MU $k$ to FS $s$ and $f_{\sf{freq}}=2.5\:GHz$. 
The noise power is set equal to $\sigma^2=10^{-13} \; W$.
In the simulation, we choose $\beta_{k,s}$ equal to $1$ for all MUs, $W_s=4$~MHz (except for Fig.~\ref{fig-Energy_W}), and $d_{s,m} = 2$~Mbits (except for Fig.~\ref{fig-Offload}). 
Each MU needs to execute a task with data size of $20$ Mbits ($D_{k,s}=20Mbit$) within a duration of $100$~ms 
($\bar{T}_{k,s}=100ms, \forall (k,s)$) (except Fig.~\ref{fig-E_Time}). 
The CPU computation capacity for each MU ($\bar{F}_{k,s}$) is randomly selected from the set $\{0.3, 0.4,..., 0.7\}$~GHz and the local computing energy per cycle is $\alpha_{k,s}=10^{-26}$~J/cycle. The number of CPU cycles per data bit  is set $c_{k,s} \in \left[ 500,1500 \right]$ cycles/bit. The computation capacities of four FSs are chosen as $\{1.7, 3.6, 3.8, 4.5\}$ GHz.
\begin{figure}[!t]
\centering
\includegraphics[width=70mm]{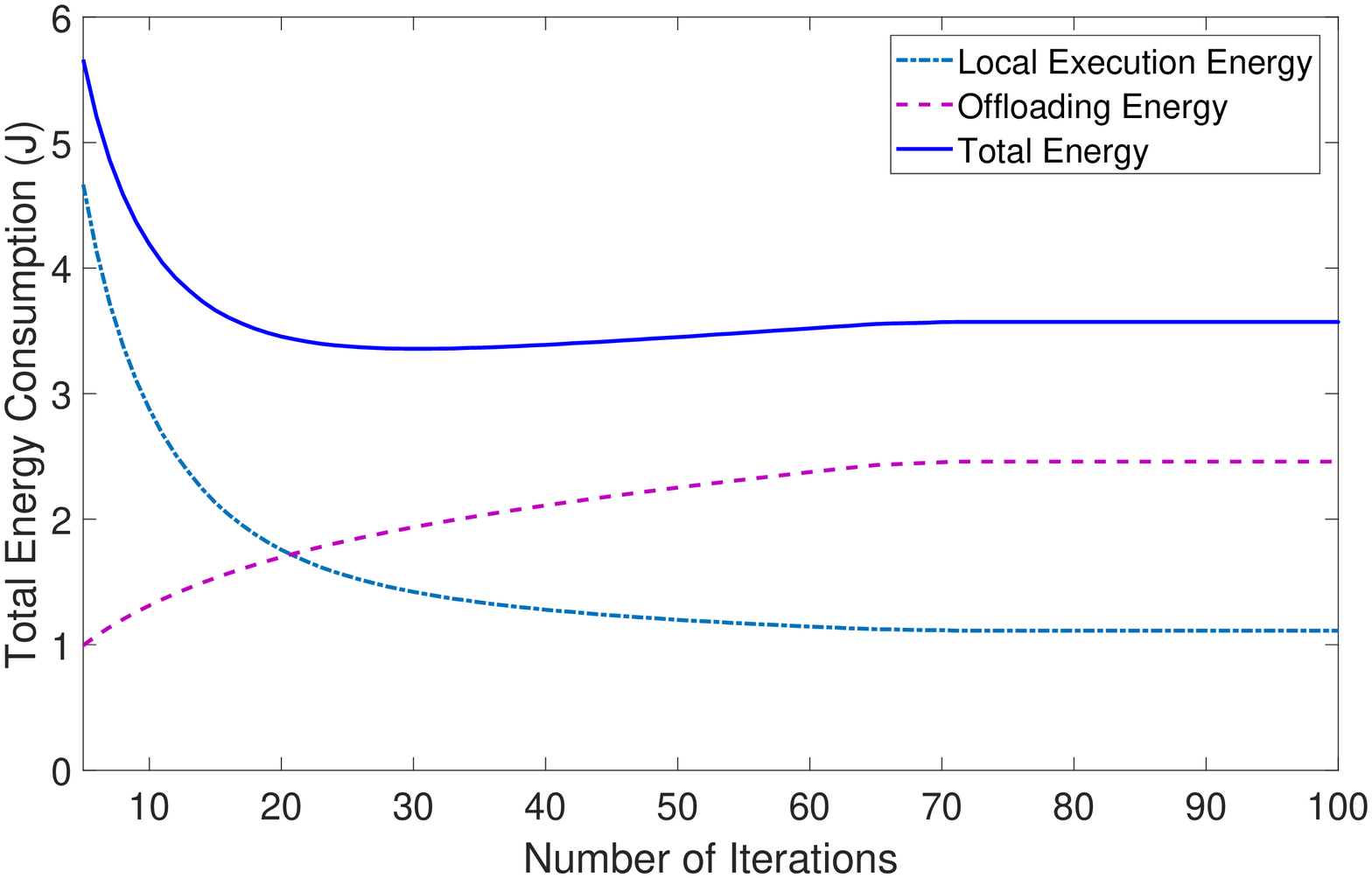}
\caption{Variations of energy consumption over iterations.}
\label{fig-Convergence}
\vspace{-3mm}
\end{figure}

First, we examine the convergence of our proposed optimal algorithm by showing the variations of total energy consumption (TEC), the energy of local execution (ELE), and offloading energy (OE) of all MUs over iterations by using Algorithm~\ref{alg:gms1} with $\rho = 100$ 
in Fig.~\ref{fig-Convergence}.
In this simulation, we consider the cooperation among the FSs under the full mesh backhaul topology as shown in Fig.~\ref{fig-MultiFog}-(a).
It can be observed that the TEC converges after $70-80$ iterations which confirms the convergence of Algorithm~\ref{alg:gms1}.
In addition, the ELE decreases while the TE increases over iterations before becoming saturated.

In Fig.~\ref{fig-Energy_W}, we show the benefit of cooperation among the FSs where the TEC versus the frequency bandwidth ($W$) is plotted for three schemes, Algorithm~\ref{alg:gms1} with full cooperation (Full Mesh) and no-cooperation, and Algorithm~\ref{alg:gms2} with full cooperation. 
As can be observed, the TEC achieved by our proposed optimal algorithm with full cooperation is much lower than those due to 
the schemes without FSs' cooperation and the greedy algorithm. 
In addition, the TEC achieved by all algorithms decrease as $W$ increases, which demonstrates
 the impact of wireless links capacity on the CO design.

Then, we illustrate the variations of TEC achieved by Algorithm~\ref{alg:gms1} with different values of allowable
 task latency ($\bar{T}_{k,s} = T, \forall (k,s)$) and application data size ($\bar{D}_{k,s} = D, \forall (k,s)$) in Fig.~\ref{fig-E_Time}.
As can be seen from this figure, the TEC decreases when $T_{k,s}$ increases while the TEC  increases when $\bar{D}_{k,s}$ becomes
larger. This is because the larger value of $T_{k,s}$ gives more freedom for the computation offloading and the larger data size 
requires higher energy for data transmission and task execution.

The performance of our proposed algorithm under different backhaul topologies is examined in Fig.~\ref{fig-E_MU} where
the TEC achieved by employing Algorithm~\ref{alg:gms1} versus the number of MUs is plotted in four backhaul network topologies: Full Mesh, Ring, Star-Max (the most powerful FS is the center), and Star-Min (the weakest FS is the center). This figure
confirms that the Full-Mesh backhaul network results in the lowest TEC while Star-Max scenario achieves better
performance than the Ring scenario, and the Star-Max is the worst.
The figure also shows that the TEC achieved by all schemes increases with the number of MUs as expected.

Finally, we demonstrate the impact of backhaul capacity on the number of bits offloaded to FSs with and without cooperation 
among the FSs in Fig.~\ref{fig-Offload}. In this simulation, we set $d_{s,m} = d, \forall (s,m)$.
As can be seen, the cooperation topologies encourage MUs to offload more data to FSs than the no-cooperation one does. 
In particular, the number of offloaded bits in Full-Mesh scheme is the highest, that number in Ring scheme is slightly smaller than that due to Star-Max scheme while the Star-Min scheme gains the lowest number of offloaded bits among all cooperation schemes.
Considering the results in both Figs.~\ref{fig-E_MU} and \ref{fig-Offload}, one interestingly confirms that the schemes having the higher number of offloaded bits can save more TEC.
In addition, the amount of offloaded data in each scheme tends to increase as the backhaul capacity increases but it 
becomes saturated when $d$ is sufficiently large.

\section{Conclusion} \label{sec:cls}

In this paper, we have considered the joint CO and RA design for backhaul limited cooperative MEC systems.
This design aims to minimize the weighted sum of energy consumption of all MUs.
We have developed the optimal algorithm using the convexification and duality methods and have described the greedy algorithm. 
Numerical results have confirmed the great performance gains of the proposed design compared 
to offloading design without cooperation among FSs and the greedy design.

%


\bibliographystyle{IEEEtran}
\bibliography{bib/refs}


\end{document}